# GRID MONITORING FOR EFFICIENT FLEXIBILITY PROVISION IN DISTRIBUTION GRIDS

*Ankur MAJUMDAR, Sotirios DIMITRAKOPOULOS, Omid ALIZADEH-MOUSAVI*

*DEPsys, Puidoux, Switzerland*
ankur.majumdar@depsys.com

## Abstract

The increased flexibility requirement needs a flexibility market at the distribution grid level operated by the distribution system operators (DSOs) to resolve challenges to ensure secure operation and, to integrate new renewable productions or loads in the grid. Therefore, the network visibility and monitoring are paramount to distribution network operation. This paper presents a methodology demonstrating the value distribution grid monitoring can bring for the realisation of a local flexibility market. This paper further illustrates the reduction of costs of operation of a DSO with the help of grid monitoring and local flexibility market while maintaining a secure and reliable operation. The methodology has been applied on a real 35 node MV network of a Swiss DSO with several GridEye measurement devices. The performance in terms of losses and voltage and flow violation costs is compared with sub-optimal operation without monitoring.

## 1 Introduction

With the increased penetration of distributed renewable generation and electrification of consumption, the modern electricity system is faced with a challenge to maintain security, reliability and quality of supply. A significant proportion of renewable generations are small-scale, intermittent and connected to the medium voltage (MV) and low voltage (LV) distribution networks. Therefore, the flexibility of the electricity system is at stake. Accordingly, there is increasing push from regulators to the DSOs to procure and activate flexibility from local distributed energy resources (DERs) such as, photovoltaics (PVs), electric vehicles (EVs), energy storages, and controllable loads [1],[2].

Moreover, the DSOs are digitalising the distribution grid with the integration of sensors, communication and controls. This has created immense opportunities for the DSOs to improve the grid operation. Technologies with monitoring, forecasting and control functionalities can provide the DSOs with a platform to access and activate the local flexibilities and to ensure that the distributed resources will not cause any violations in the grid operational limits.

In line with the Clean energy Package 4 (CEP4), it is recommended that flexibilities are procured and provided through market mechanisms and there is a level-playing field for every distributed and flexible resource in the market. It is argued that the local problems of the DSO such as, voltage violation and congestion should be resolved effectively by acquiring and activating flexibilities at the local market level [3],[4]. Hence, there is increasing focus on developing local flexibility market in many countries. At this juncture, improved grid monitoring, forecasting and the ability to communicate control set-points among different players/resources at the MV and LV levels play crucial roles.

The distribution networks have been traditionally operated with no monitoring at the LV level and limited to almost no monitoring at the MV level. However, with more DER installations at the MV and LV grids and requirement of flexibilities from the local resources means that DSOs require more monitoring information of their grid both at the MV and LV levels.

Smart meters are being or in the process of being rolled out in many countries in Europe and elsewhere initially for billing purposes [5],[6]. Many countries have smart meter rolling out as a priority in their energy strategies. Many DSOs have also started pilots to improve the network visibility with smart meters. However, the privacy and access to smart meter data is a challenge and the accuracy of smart meter information is a concern for grid analysis purposes. Besides, according to current practice, smart meters transfer 15-minutes data once per day as the DSOs would have to bear the high cost associated with faster data transfer and higher resolution data storage for all end customers. Hence, there is a requirement for grid monitoring devices, which are plug-and-play devices and provide real-time measurement.

This paper discusses the roles of distribution grid monitoring technologies for achieving a market operated by the DSO for procuring local flexibilities. The study illustrates how monitoring at the LV side of the MV/LV transformer can better realise the flexibility market for the DSO, and thus reduce their costs of operation and ensure the security and quality of supply.

## 2 Methodology – Local Flexibility Market

The modern distribution networks are in general not monitored at the MV and LV levels. At the MV level, monitoring is only available at the HV/MV substation and sometimes it may or may not be available at the departure of every MV feeder. In this scenario, the monitored values at every MV feeder are allocated to the MV/LV transformers according to the rating of the transformers. This may lead to voltage violation or congestion and to increased load scheduling costs at the MV/LV transformers. Therefore, it is difficult to realise the flexibility potential at the MV and LV level. The current distribution operation is based on load disaggregation (where the loads are allocated based on the transformer kVA rating), on-load tap changer operation of the HV/MV transformer, and the grid security actions are dealt with in a sub-optimal way [7],[8]. Nonetheless, with more distributed resources connected at the MV and LV grids, the DSOs require more visibility of their grid and an optimal operation of their grid and their assets (optimal power flow).

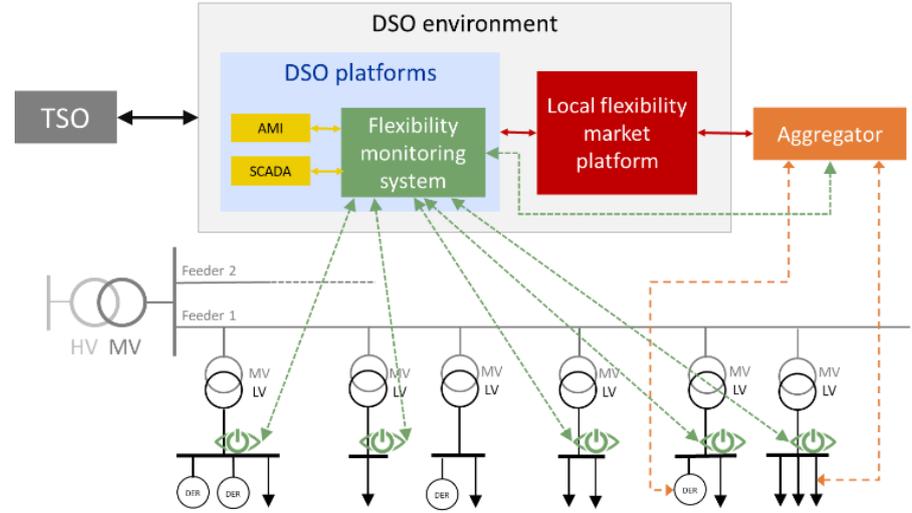

Fig. 1. The proposed flexibility platform for DSO and the roles of monitoring technology.

The proposed method in this paper consists of developing the local flexibility market, as in Fig. 1, maintained by DSO. The figure shows how the flexibility monitoring system communicates with the platforms in DSO environment and the aggregator platform for an efficient realisation of the local flexibility market. The approach will be realised by real-time, accurate and detailed information of the grid status from the grid monitoring devices. Furthermore, this will be achieved through synergies with aggregator (better optimisation of its portfolio with monitoring information provided by the flexibility monitoring system), DSO and TSO through increased data and information exchange.

The proposed flexibility market is based on multi-objective optimisation minimising the DSO operational costs (maintaining voltage levels, congestion management, losses reduction), DSO costs to maintain TSO requirements (interconnection flow and voltage level) and the DSO costs for scheduling and control of the distributed resources. This is applied on a real grid of a Swiss DSO equipped with GridEye monitoring devices with accurate 10-minute interval data.

In order to integrate flexibility in their portfolio and to facilitate the energy transition, the DSOs will require to play a more proactive role in its operation and have four main objectives:

1. Reducing the security costs – to ensure secure operation, the DSOs will have to prevent the voltage violations and the congestions.
2. Reducing the operational costs – to ensure optimal operation of their assets, DSOs will have to minimise the losses or reactive power flow in their network.
3. Reducing the scheduling costs – in order to ensure cost-efficient grid operation, DSOs will require to minimise the scheduling costs of controllable DERs, curtailment of load and/or DERs
4. Reducing the costs for maintaining TSO set-points – DSOs will require to minimise the deviation from the agreed active and reactive power schedules with the TSO. Otherwise, the DSOs pay penalties to the TSO for the deviation in the agreed schedules.

The formulation of the problem is an optimal power flow described below.

$$min\ w_I \sum_{lines} r_{ij}|I_{ij}|^2 + w_V \sum_{nodes} V_{dev}$$
$$+w_{Ilim} \sum_{lines} |I_{ij}|^2_{dev} + w_p \sum_{nodes} p_i + w_q \sum_{nodes} q_i \quad (1)$$

s.t. $\sum_{adj} P_{ij} + p_i = p_i^g - p_i^c \quad \forall\ node, \forall\ adj\ lines \quad (2)$

$\sum_{adj} Q_{ij} + q_i = q_i^g - q_i^c \quad \forall\ node, \forall\ adj\ lines \quad (3)$

$|V_j|^2 = |V_i|^2 - 2r_{ij}P_{ij} - 2x_{ij}Q_{ij} + (r_{ij}^2 + x_{ij}^2)|I_{ij}|^2 \forall line (4)$

where, $V_{dev} = 0$, if $V^{min} \leq V_i \leq V^{max}$ or, $V_{dev} = |V_i|^2 - V_{lim}^2$, if $V_i < V^{min}$ or $V_i > V^{max}$ and, $|I_{ij}|^2_{dev} = 0$, if $|I_{ij}| \leq I_{max}$ or $|I_{ij}|^2_{dev} = |I_{ij}|^2 - |I_{ij}|^2_{max}$, if $|I_{ij}| > I_{max}$, representing the penalty costs for voltage and flow violation by piece-wise linear functions. $r_{ij}$ and $x_{ij}$ represent the line resistance and reactance respectively for line $ij$. Eq. (1) is the objective function including losses, voltage and line flow violation costs, and load curtailment and/or DER scheduling costs with their associated weighting factors, respectively given by $w_I$, $w_V$, $w_{Ilim}$, $w_p$ and $w_q$. This is a multi-objective problem subject to equality constraints for nodal active and reactive power balances (2)-(3) and network related constraints (4). Since the case for study here is a radial MV distribution network, shown in Fig. 2, the equations are formulated as linear DistFlow equations with $v_i = |V_i|^2$ and $l_{ij} = |I_{ij}|^2$ as variables [9] and with necessary convex relaxation [10].

## 3 Results

The use case objective is to define a local flexibility market platform and to valorise the DSO operational costs in three different cases

- **Case 0:** No market with monitoring at secondary of MV/LV transformer
- **Case 1:** Market for flexibility with no monitoring at secondary of MV/LV transformer
- **Case 2:** Market for flexibility with monitoring at secondary of MV/LV transformer

**Case 0** represents the distribution system operation based on real-time monitoring at the secondary of the MV/LV level but with no market operation. Hence, the DSO set-points are decided based on a power flow of the MV network with measurements at the MV/LV transformer nodes. **Case 1** represents the system operation with no monitoring information but with a market for flexibility for the DSO. The loads at each MV/LV transformer are allocated based on the kVA rating of the transformers with the same profile as the MV feeder they are connected. The case is simulated by running an optimal power flow (OPF) with the allocated loads. A power flow is performed with the OPF set points to calculate the realised security costs and losses. Finally, **Case 2** represents the DSO operation with a flexibility market with real-time monitoring information at MV/LV transformer nodes. This case is also simulated by performing an OPF with real measurement data at the secondary of the MV/LV transformer to generate the optimal set points and calculate the security costs and losses.

Fig. 2 shows a schematic diagram of a radial MV network of a Swiss DSO with 34 lines and 35 MV nodes. The figure also shows the location of grid monitoring devices, GridEye, on the secondary of the distribution (MV/LV) transformers providing the synchronised measurement data for voltages, active and reactive powers for each MV/LV transformer. The dataset is 10-minutes measurements of active and reactive powers.

Fig. 3 shows a representative daily feeder level active power profiles for the two feeders of the distribution network. When the DSO has no monitoring (**Case 1**), the loads are allocated to the MV/LV transformers based on their kVA ratings with the same profile as the MV feeder they are connected to.

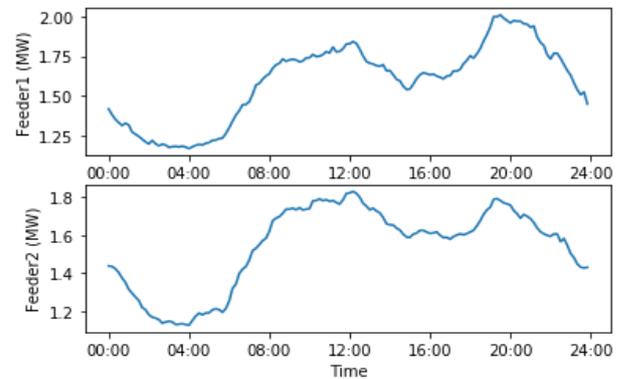

Fig. 3. The load profiles of two feeders from the HV/MV substation for a representative day.

The methodology is illustrated for two scenarios. Scenario 1 represents the normal loading condition for Feeder1 and Feeder2. Scenario 2 represents an additional load condition representing electrification of transport, heating and cooling.

For Scenario 1, Table 1 shows a comparison of the sum of the technical losses in kWh over 24 hours. The values show that there is a 9.1% decrease in losses in Case 2 compared to 8.9% decrease in Case 1 in relation to Case 0. It is to be noted that there is no security violation (voltage/flow) for this scenario.

Fig. 4 shows the technical losses and security violation costs in CHF over 24 hours. This scenario shows that with the monitoring at the secondary of MV/LV transformers it is possible to reduce the technical losses of the DSO while operating a flexibility market. The optimal power flow in **Case 2** generates the optimal set points compared to non-optimal set-points in **Case 1** and **Case 0**.

In Scenario 2, additional loads of 300 kW representing electrification of transport, heating and cooling are added at each node. In this scenario, due to additional loads there are additional flows in the lines in two feeders. Among them the two lines on Feeder1 are congested with flows more than their security limits.

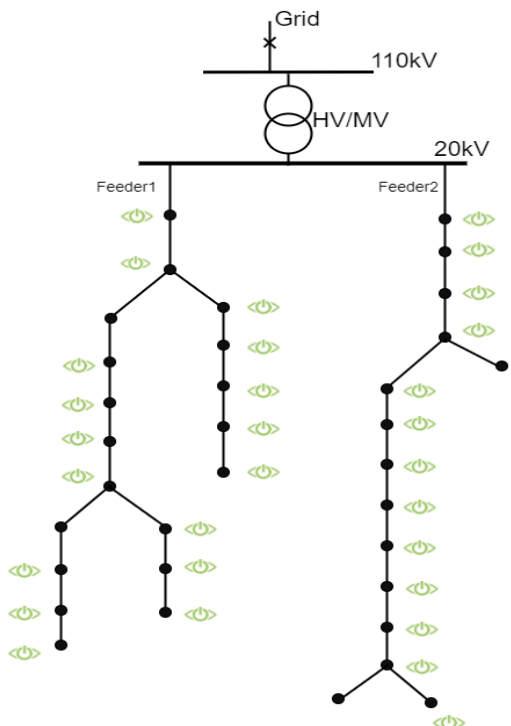

Fig. 2. A MV distribution network of a Swiss DSO equipped with monitoring devices.

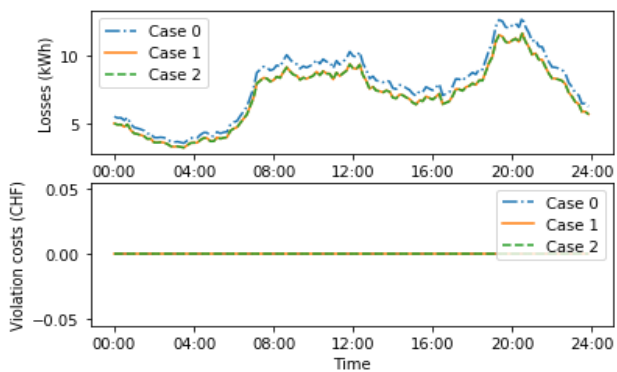

Fig. 4. Losses and security violation costs for Scenario 1.

Table 1. DSO technical losses in kWh and security violation costs in CHF for Scenario 1.

|  | Case 0 | Case 1 | Case 2 |
| --- | --- | --- | --- |
| Losses | 1121.1 | 1021.1 | 1019.2 |
| Security violation | 0 | 0 | 0 |

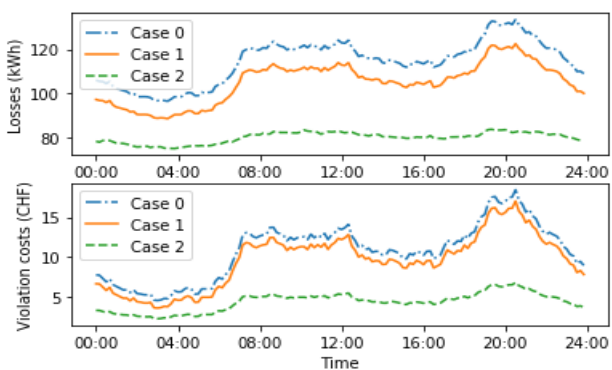

Fig. 5. Losses and security violation costs for Scenario 2.

Table 2. DSO technical losses (kWh) and security violation (CHF) over 24 hours for Scenario 2 with 300 kW additional loads.

|  | Case 0 | Case 1 | Case 2 |
| --- | --- | --- | --- |
| Losses | 16476.90 | 15124.40 | 11519.70 |
| Security violation | 1552.62 | 1382.52 | 632.81 |

Table 2 shows the sum of the losses in kWh and security violation costs in CHF over 24 hours for the three cases. **Case 2**, with optimal set points has less violation costs (59.2% decrease from **Case 0**) and less losses (30.1% decrease from **Case 0**) compared to **Case 1** with 10.9% decrease in violation costs from **Case 0** and 8.2% decrease in losses from **Case 0**. Fig. 5 further illustrates the evolution of losses in kWh and security violation costs in CHF over 24 hours. This clearly demonstrates the value of monitoring in a flexibility market platform with minimum losses and minimum violation costs.

## 4 Conclusion

The idea of monitoring and marketplace for flexibility at MV and LV grids will be beneficial for all the grid operators. This will effectively reduce the grid operation costs, facilitate DSOs to be better prepared for their energy scheduling and decrease the rescheduling costs of TSO, and increase the hosting capacity for new renewable productions and electrification of consumptions. The simulation and results presented in this study indicate that monitoring at the distribution level is beneficial for reducing the security violation risks and losses but also to operate a flexibility market. In the next steps, the role of aggregators will be included to demonstrate how the aggregators can efficiently optimise their portfolio with better monitoring information. This will enable a complete realisation of the local flexibility market. Therefore, this study will provide a roadmap for achieving a flexible, decarbonised and digitalised electricity system.

## 5 Acknowledgement

This project is supported by the European Union's Horizon 2020 programme under the Marie Sklodowska-Curie grant agreement no. 840461.